 \documentclass[aps,prb,twocolumn,groupedaddress]{revtex4}
\usepackage{graphicx}

\begin{document}

\title{Theoretical study of magnetic properties and hyperfine interactions in
$\sigma$-FeV alloys}

\author{J. Cieslak}
\email[Corresponding author: ]{cieslak@novell.ftj.agh.edu.pl}
\author{J. Tobola}
\author{S. M. Dubiel}
\affiliation{Faculty of Physics and Applied Computer Science,
AGH University of Science and Technology, al. Mickiewicza 30, 30-059 Krakow, Poland}

\date{\today}

\begin{abstract}

Electronic structure Korringa-Kohn-Rostoker calculations for the $\sigma$-phase in Fe$_{100-x}$V$_x$ were performed in
compositional range of its occurance ($\sim 34 \le x \le \sim 65$). Fe and V magnetic moments and hyperfine fields were determined
for five inequivalent lattice sites in two models of magnetic structure, namely ferromagnetic FM and so-called anti-parallel one, APM,
dominated by antiferromagnetic coupling.
The average magnetic moments calculated for FM state overestimate the experimental data, whereas the corresponding quantities computed
for APM state, underestimate them. Such a behavior remains in line with total energy values being similar for both models.
The calculations showed that both average magnetic moments and hyperfine fields (on Fe and V atoms) vary with a number of Fe atoms in
the nearest neighbor shell, $NN_{Fe}$, starting from critical values of $NN_{Fe}$, characteristic of each site.

The calculated hyperfine fields for Fe and V showed an important role of valence contributions, being strongly dependent on
local magnetic moments arrangements. In the case of Fe, the computed hyperfine fields are overestimated with respect to the Mossbauer
data, while the corresponding values for V are underestimated with respect to the NMR results. However, the linear correlation
between average magnetic moment and hyperfine fields, observed experimentally in FeV $\sigma$-phase, can be well reproduced
when combining theoretical results for the two above-mentioned magnetic structure models.

\end{abstract}


\maketitle

\section{Introduction}

Among the so-called Frank-Kasper phases that can be geometrically described in terms of the basic
coordination polyhedron with coordination number equal to 12, 14, 15 and 16 \cite{Frank58},
the $\sigma$-phase that can occur
in transition-metal alloy systems, is known as the one without definite stoichiometric
composition.  Consequently, it can exist in different composition ranges and, therefore, its
physical properties can be tailored by changing its constitutional elements, and within a given
constitution, by changing chemical composition.  Among about 50 examples of binary
$\sigma$-phases that have been reported so far in the literature, only two viz.  $\sigma$-FeCr
and $\sigma$-FeV are known to have well evidenced magnetic properties
\cite{Parsons60, Hall66, Read66, Read68, Cieslak04, Cieslak08a, Cieslak09},
though a knowledge
of the $\sigma$-phase magnetism remains not complete.  In particular, neither values of the
magnetic moments, $\mu$, on particular lattice sites nor a coupling between them is known.  A
lack of stoichiometry, leading to a huge number of different atomic configurations, combined with
a weak magnetism i.e.  small values of $\mu$ makes it very difficult for microscopic methods like
M\"ossbauer spectroscopy or nuclear magnetic resonance to uniquely determine spin-densities at
the level of the sublattices.  Non-availability of big enough single-crystals of the
$\sigma$-phase prevents the neutron diffraction techniques to be used for that purpose.  In these
circumstances, performing theoretical calculations aimed at determining electronic and magnetic
structures of the phase is highly needed and helpful.

The present paper reports such calculations for the Fe$_{100-x}$V$_x$ system where the
$\sigma$-phase exists in a wide range of composition ($\sim 34 \le x \le \sim 65$), and its
magnetic properties show a strong $x$-dependence.  For example, the Curie temperature may vary
between 0 K, for $x\approx 65$, and $\sim$320 K, for $x\approx 34$ \cite{Cieslak09}.
This gives a unique chance for testing the calculations which were already
successfully applied for similar purpose in the $\sigma$-FeCr alloy system \cite{Cieslak10b}
where the phase exists in a much narrower compositional and temperature ranges.

\begin{figure*}[tb]
\includegraphics[width=.99\textwidth]{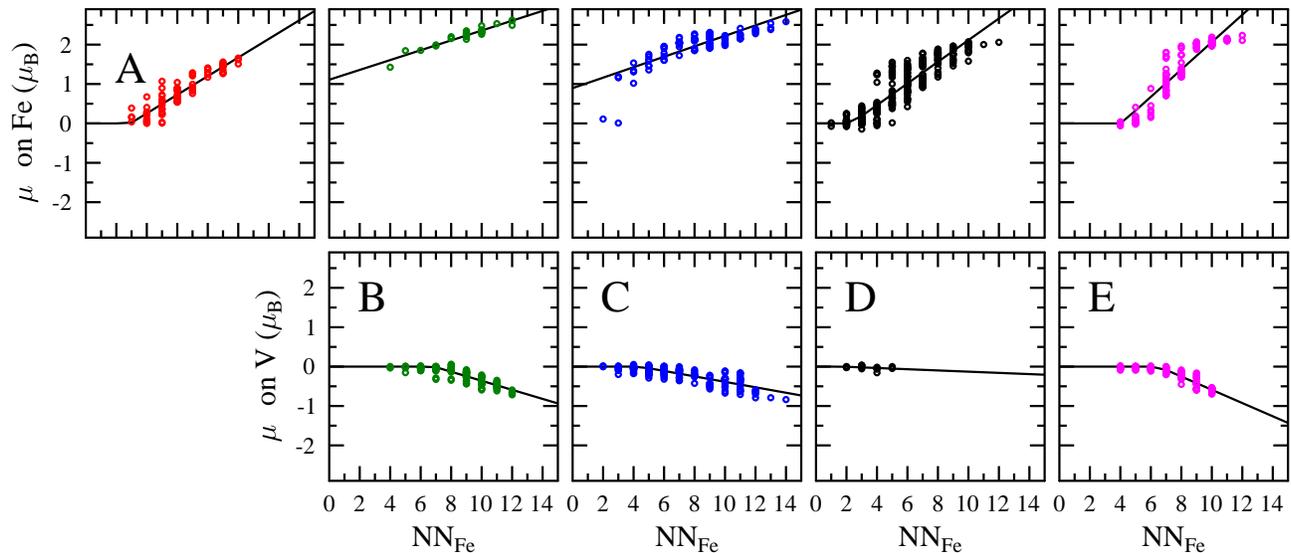}
\caption{(Online color)
Magnetic moments of Fe and Cr atoms for five crystallographic sites versus the number of
$NN_{Fe}$ atoms as obtained with the FM model of the magnetic ordering.  Solid lines stand for
the best fits to the data.
}
\label{fig1}
\end{figure*}

\begin{figure*}[bt]
\includegraphics[width=.99\textwidth]{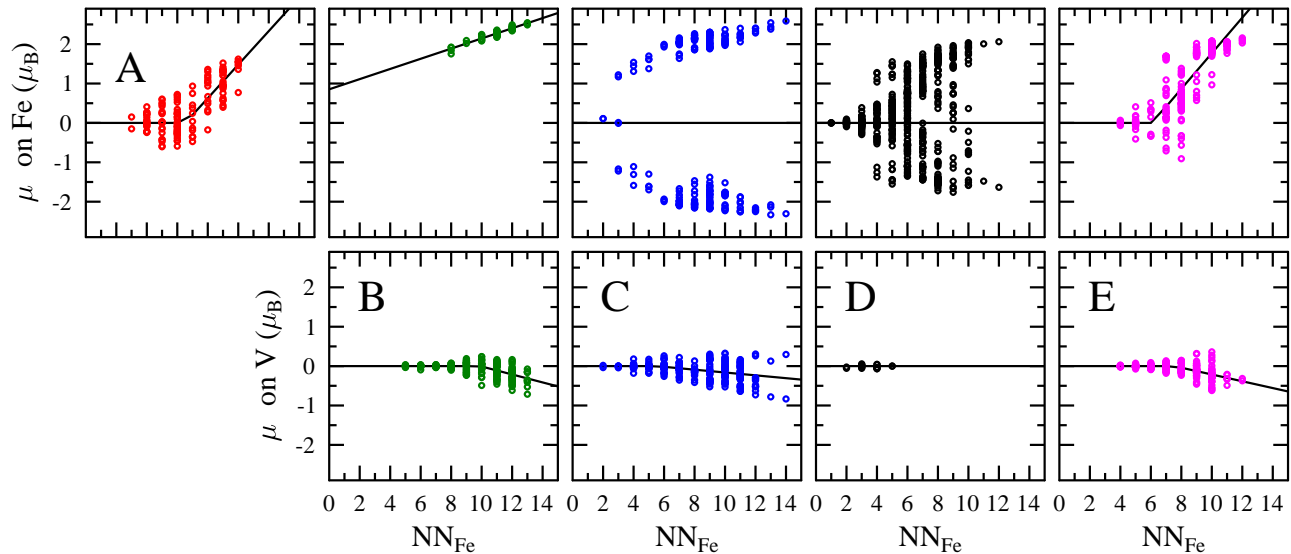}
\caption{(Online color)
Magnetic moments of Fe and Cr atoms for five crystallographic sites versus the number of
$NN_{Fe}$ atoms as obtained with the APM model of the magnetic ordering.  Solid lines stand for
the best fits to the data.
}
\label{fig2}
\end{figure*}

\section{Computational Details}

Electronic structure calculations for $\sigma-$FeV compounds have been performed
using the charge and spin self-consistent Korringa-Kohn-Rostoker (KKR) method. The
muffin-tin crystal potential has been employed within the local
density approximation (LDA) framework, where the von Barth-Hedin formula
\cite{Barth72} for the exchange-correlation part was used. The self-consistent
crystal potentials were converged below 0.1 mRy and charges below $10^{-3}~e$.
The spin-polarized densities of states were computed employing the tetrahedron {\bf k}-space
integration technique. The local magnetic moments (of Fe and V) inside each muffin-tin spheres
were determined. Accordingly, the Fermi contact hyperfine fields ($H_{hf}$) resulted from
the spin densities extrapolated to $r=0$, $\rho(0)$, with the use of the well-known
non-relativistic formula $H_{hf}= {8 \pi\over3} \mu_B \rho(0) \approx 524 \rho(0) $ in kOe
\cite{Blugel87}. The KKR calculations allowed to determine core ($H_{core}$) and valence
($H_{val} = H_{hf} - H_{core}$) terms, neglecting however dipolar and orbital contributions to the total hyperfine
field. It is worth noting that electronic structure calculations have been done for ordered
approximants of the disordered $\sigma-$FeV alloys, and for that purpose the symmetry
of the unit cell was lowered to a simple tetragonal one. In practice,
the tetragonal unit cell (space group P$4_2$/mnm) and atomic positions were
unchanged but variable occupancy made all 30 atomic positions crystallographically
nonequivalent (space group P$1$) being next occupied exclusively either by Fe or V atom.
Calculations have been performed for two $\sigma$-FeV alloys with different compositions,
namely Fe$_{20}$V$_{10}$ and Fe$_{12}$V$_{18}$.
They are equivalent to the formula Fe$_{100-x}$V$_{x}$ with $x=33.3$ and $x=60.0$, respectively.
The chosen compositions correspond fairly well to border concentrations of the $\sigma$-phase existance in
the Fe-V alloy system \cite{Kubaschewski82}. Fe and V atoms were distributed over five
inequivalent sublattices of the tetragonal unit cell in close agreement with experimentally determined occupancies
\cite{Cieslak08c}.  Actually, the used here atomic configurations in the unit cell remained strictly
the same as those considered previously in electronic structure calculations of the $\sigma$-FeV
phase in a paramagnetic state \cite{Cieslak10a}.

\section{Results and Discussion}

\subsection{Magnetic Moments}

In order to study magnetic properties from first principles calculations, two collinear models of
a possible magnetic ordering were taken into account:  (i) parallel alignment of local magnetic moments,
called here ferromagnetic, FM, as derived directly from spin-polarized calculations, and (ii)
antiparallel alignment, called here APM. In the latter, Fe/V atoms in selected sublattices were
initially coupled antiferromagnetically according to predictions of a symmetry analysis as given
in details elsewhere \cite{Cieslak10b}.  The magnetic moments and hyperfine fields, $H_{hf}$,
corresponding to both Fe and V atoms distributed over the five sublattices were a subject of a
further analysis.  Their non-zero values were found for both Fe and V atoms.  As a rule, the
magnetic moments (and the hyperfine fields) on V atoms were found to be much smaller, and,
generally, polarized opposite, to those on Fe ones.

The calculated values of $\mu$ and $|H_{hf}|$ for each sublattice were analyzed versus a number of Fe
atoms in the nearest neighborhood, $NN_{Fe}$. Such approach combining the results obtained for
the two border concentrations, shows that the two sets of the computed data are complementary, and
they cover the expected range of $NN_{Fe}$ for all sublattices.

%comment
%
%In all considered cases, no changes of the nature of the magnetic properties, which would be caused by the concentration of the alloy
%chosen for calculations were observed. Accordingly, at all further analysis of $\mu(NN_{Fe})$ and $<|H|>(NN_{Fe})$ the concentration
%was not taken into account. In other words, we have considered only the effect of the $NN_{Fe}$-numbers without taking into account
%concentration for which these figures have been designated.

A detailed analysis of the $\mu(NN_{Fe})$-dependence of the magnetic quantities for each of the
sublattices shows that this relationship can be split, in most cases, into two regions showing a
linear character.  In the investigated $\sigma$-FeV system, an average magnetic moment per an
atom, $<\mu>$, on a given sublattice remains zero up to a critical value of
$NN_{Fe}={NN_{crit}}$, and for $NN_{Fe}\ge{NN_{crit}}$ it varies linearly in accord with the
equation

\begin{equation}
  <\mu>=\left\{\begin{array}{lll}
                   a(NN_{Fe}-NN_{crit})   &   \mbox{for} &  NN_{Fe}>NN_{crit} \\
                                      0   &   \mbox{for} &  NN_{Fe}\le NN_{crit} \\
                  \end{array}
          \right.
\end{equation}

A similiar dependence on $NN_{Fe}$ was also obtained for the average $|H_{hf}|$, $<|H_{hf}|>(NN_{Fe})$, for
all sublattices, and both considered models of the magnetic ordering (FM and APM).

Noteworthy, a similar discontinous $\mu(NN_{Fe})$-dependence was already reported in the analysis
of the atomic moment in {\it bcc} Fe-V alloys \cite{Shiga78}, and also used in calculations
of the hyperfine field distribution curves for nanocrystalline {\it bcc} Fe-V alloys \cite{Ziller01}.  The
critical value of $NN_{Fe}$ as found by these authors was rather small, $NN_{crit}=3$, when
accounting for the magnetic moments localized on Fe-atoms.  In the present case of the
$\sigma$-phase, $NN_{crit}$ values and slopes $a$ of the lines were found by a fitting the data
calculated for Fe and V atoms located on five various sublattices, separatelly.  The obtained
$NN_{crit}$-values in that way do not exceed 6, as can be clearly seen in Figs.~\ref{fig1} and~\ref{fig2}.
It is worth noting that negative values of $NN_{crit}$, as found in several
cases, mean that the magnetic moments do not vanish for all $NN_{Fe}$-values, as derived from Eq.
1.

\begin{figure}[tb]
\includegraphics[width=.49\textwidth]{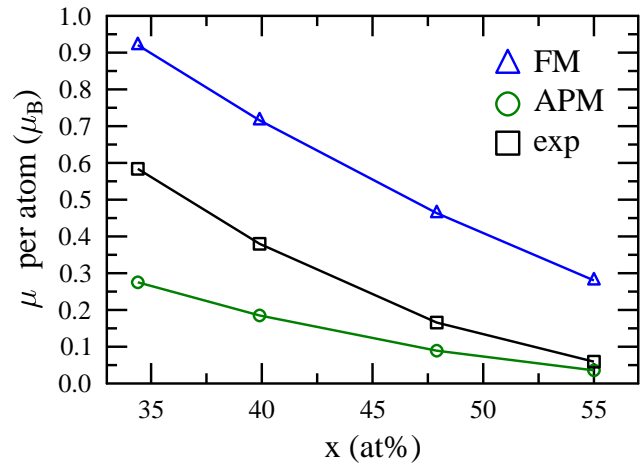}
\caption{(Online color)
Average magnetic moments as calculated in FM (triangles) and APM (circles) models, as well as
experimentally determined data (squares) \cite{Cieslak09}.  Solid lines are for guide eye only.
}
\label{fig3}
\end{figure}

The calculated magnetic moments for both types of atoms located on all crystallographic positions
combined with calculated probabilities of these position occupancies allowed to determine the average values
of the magnetic moments, $<\mu>$, for each sublattice and for any concentration of vanadium.  In
the same way, one could compute the average magnetic moment per atom (per unit cell).  In practice,
such calculations were performed for four selected concentrations that corresponded to the samples
on which the magnetic measurements were performed.  A comparison between the calculated and
experimentally determined quantities is displayed in Fig.~\ref{fig3}.  It can be clearly seen that the
values of $<\mu>$ calculated applying the FM model are systematically larger than those found
experimentally, $<\mu_{exp}>$.  The difference between the former and the latter changes from
about $\sim 0.35\mu_B$ per atom for $x=34.4$ to $\sim0.25 \mu_B$ for $x=55.0$.  However, the
overall dependence of $<\mu>$ on the vanadium content follows well that found experimentally.
On the other hand, the $<\mu>$-values derived from the APM model are underestimated in comparison to
those of $<\mu_{exp}>$ but the agreement bewteen theoretical and experimental values
markedly improves with increasing V content. For the vanadium richest sample, Fe$_{45}$V$_{55}$,
the measured value of $<\mu_{exp}>=0.05 \mu_B$ is practically the same as the one calculated
with the APM model, $<\mu>=0.04 \mu_B$.

\begin{figure*}[bt]
\includegraphics[width=.99\textwidth]{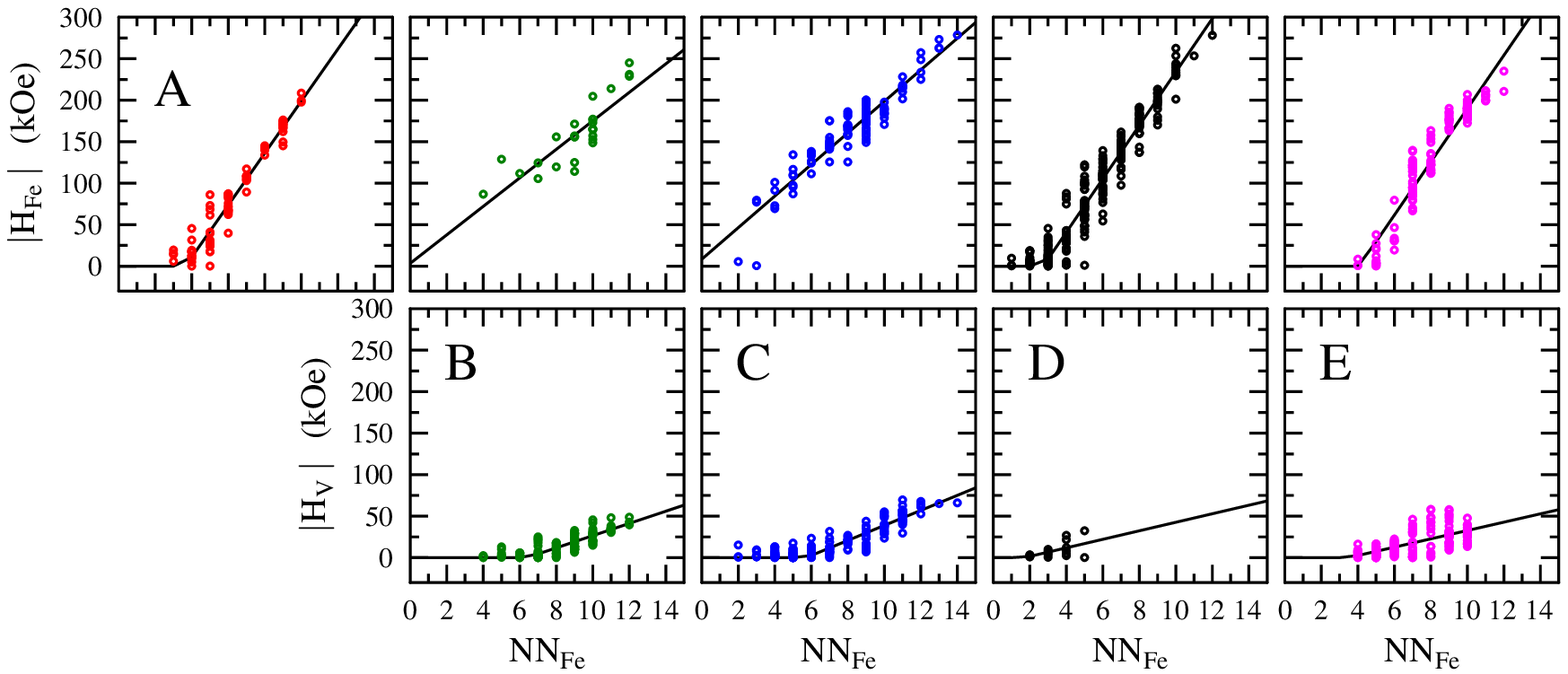}
\caption{(Online color)
$|H_{V}|$ and $|H_{Fe}|$-values for five crystallographic sites versus the number of $NN_{Fe}$
atoms as obtained with the FM model of the magnetic ordering.  Solid lines stand for the best
fits to the data.
}
\label{fig4}
\end{figure*}

\begin{figure*}[bt]
\includegraphics[width=.99\textwidth]{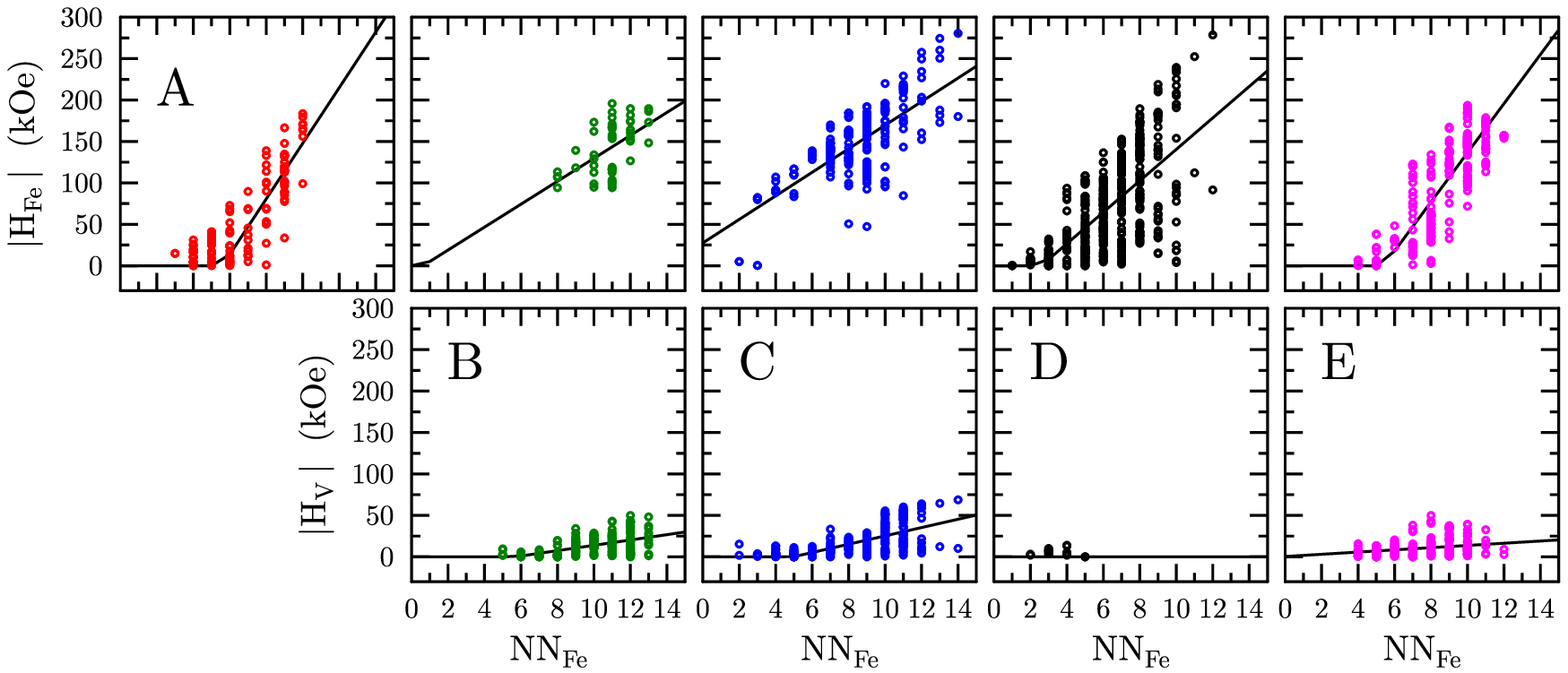}
\caption{(Online color)
$|H_{V}|$ and $|H_{Fe}|$-values for five crystallographic sites versus the number of $NN_{Fe}$
atoms as obtained with the APM model of the magnetic ordering.  Solid lines stand for the best
fits to the data.
}
\label{fig5}
\end{figure*}

\begin{figure}[bt]
\includegraphics[width=.49\textwidth]{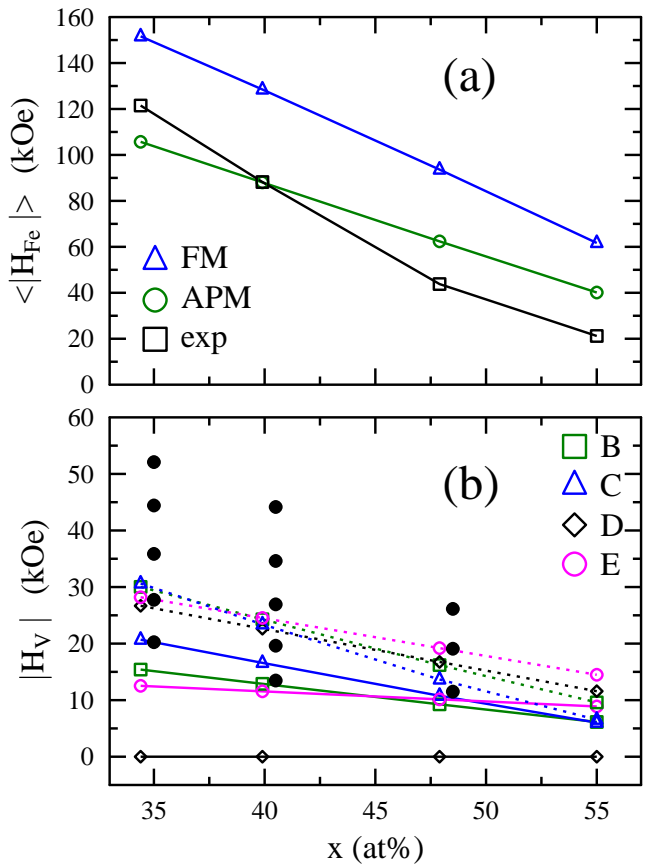}
\caption{(Online color)
(a) Average absolute values of hyperfine fields on Fe atoms as calculated in FM (triangles) and
APM (circles) models, as well as experimentally determined data (squares) \cite{Cieslak09}. Solid lines are to
guide the eye only.
(b) Average absolute values of hyperfine fields on vanadium atoms for five crystallographic sites
as calculated in FM (dashed lines) and APM (solid lines) models, as well as experimentally
determined data (circles) \cite{Dubiel10}.
}
\label{fig6}
\end{figure}

\begin{figure*}[bt]
\includegraphics[width=.99\textwidth]{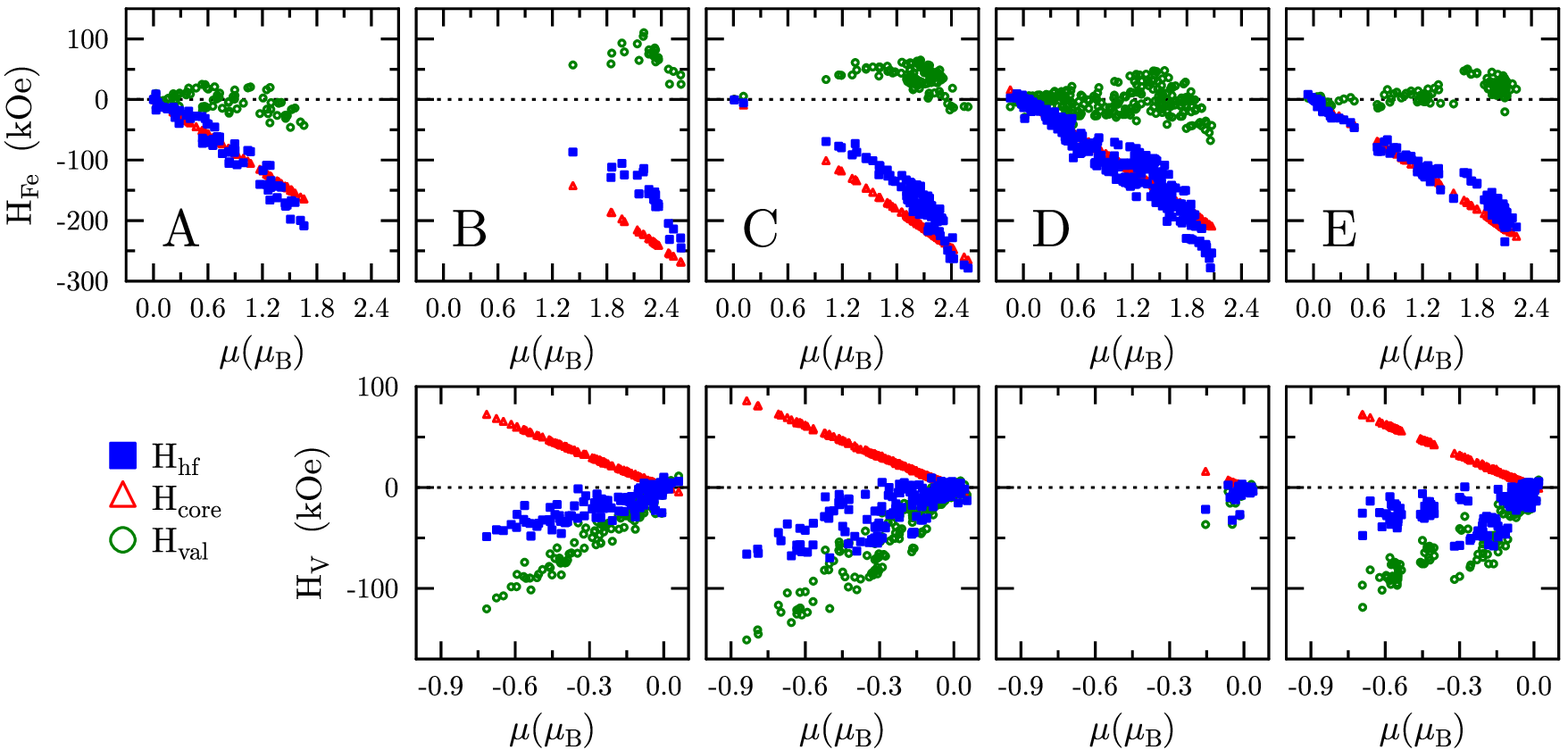}
\caption{(Online color)
$H_{hv}$, $H_{core}$ and $H_{val}$-values for five crystallographic sites versus the
magnetic moment $\mu$ as obtained with the FM model in $\sigma$-FeV phases.
}
\label{fig7}
\end{figure*}

The relationship between the measured and calculated values of $<\mu>$ can be explained in terms
of a total energy of the $\sigma$-phase, $E$, calculated for the two models of magnetic of ordering
($E_{FM}$ and $E_{APM}$). The energy difference $\Delta E = E_{APM} - E_{FM}$, was determined
using the electronic structure calculations for the two border concentrations of the $\sigma$-phase
within both FM and APM models.  In the case of the $\sigma$-Fe$_{20}$V$_{10}$
alloy $\Delta E$-value ranges from $\sim 0.7$ to $\sim 1.5$ mRy per atom, that would favour the FM model.
For the $\sigma$-Fe$_{12}$V$_{18}$ alloy the $\Delta E$-value almost vanishes (less than 0.05 mRy).
The total energy difference of the $\sigma$-phase can be compared to KKR results obtained for FM and (hypothetical)
AFM ordering in a simple $\alpha$-phase with the same Fe-V composition.
It yields the value of $\sim 15$ mRy per atom (also favoring FM ordering), being thus an order of magnitude
larger than in the former case.

Hence, the small difference between total energies computed for the FM and APM structures may
suggest that some regions of the sample have the antiparallel magnetic order (APM), whereas in
others the magnetic moments arrange mostly ferromagnetically (FM).  In other words, the magnetic
properties of the $\sigma$-FeV can be thought of as a coexistence of two types of the ordering.
The relative contribution of them should depend on the alloy composition. In view of the total
energy analysis, the contribution of the APM-like regions should be increasing with the vanadium
concentration, $x$, since the $\Delta E$ value decreases with $x$. The latter effect is also
observed experimentally in the dependence $<\mu_{exp}>(x)$ - Fig.~\ref{fig3}.

On the other hand, it should be remembered that both FM and APM models do not strictly refer to
the ferromagnetic and antiferromagnetic ordering of Fe/V magnetic moments in the
$\sigma$-phase. They should be rather treated as starting configurations, and finally converged
magnetic moment values and their orderings may differ from the initial alignments.
In consequence, magnetic moments on Fe/V atoms may vary both in magnitude and in a
mutual coupling, also due to the fact that the $\sigma$-FeV phase is a chemically
highly disordered system. In particular, it is possible that the real magnetic ordering
is not collinear. A necessary condition for the occurence of spin-canting i.e. competition of
ferromagnetic and antiferromagnetic interactions, has been here, and also in the case of the
$\sigma-FeCr$ \cite{Cieslak10b} theoretically suggested to exist.

\begin{figure}[bt]
\includegraphics[width=.49\textwidth]{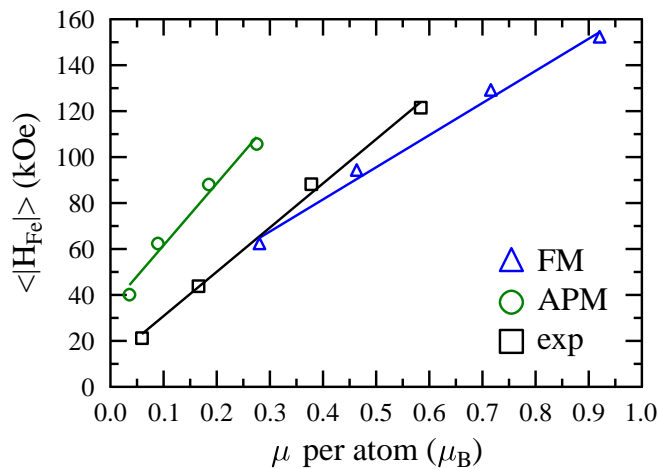}
\caption{(Online color)
Average hyperfine field on Fe-atoms, $<|H_{Fe}|>$, versus average magnetic moment per atom,
$\mu$, in the $\sigma$-FeV phase, as measured and calculated in FM and APM models.
}
\label{fig8}
\end{figure}

\subsection{Fe and V hyperfine fields}

Another quantity pertinent to magnetic materials that has been calculated is a hyperfine field,
$H_{hf}$. Although its correlation with the magnetic moment is usually non-linear and, in some
instances, even non-monotonous \cite{Dubiel09}, in the case of the $\sigma$-FeV alloys $|H_{hf}|$
and $\mu$ exhibit a perfect linear correlation \cite{Cieslak09}.  In addition, for
the $\sigma$-FeV compounds the hyperfine fields had been measured both on $^{57}$Fe and $^{51}$V
\cite{Cieslak09, Dubiel10}, so they can be used as a proper data set for testing pertinent theoretical
calculations. Those obtained presently by means of the KKR calculations for the FM and APM models are
shown in Fig.~\ref{fig4} and~\ref{fig5}, respectively.  As can be clearly seen one can distinguish two
rangies for all sublattices and occupations, namely zero-field range ($H_{hf} = 0$) for small values
of $NN_{Fe}$, and proportionality range ($H \sim NN_{Fe}$) for higher $NN_{Fe}$-values.  In order
to compare the calculated $H_{hf}$-values with the experimental
ones (M\"ossbauer and NMR data) we consider only the absolute values, $|H_{hf}|$.

Concerning the Fe-site hyperfine fields, we can make a comparison - see Fig.~\ref{fig6}a - only between
their average values, $<|H_{Fe}|>$, due to the fact that experimentally only such quantity is
available from the $^{57}$Fe-site M\"ossbauer spectroscopic measurements \cite{Cieslak09}.  As it follows
from Fig.~\ref{fig6}a, the average hyperfine field on Fe atoms as calculated in the frame of the FM model
decreases linearly with $x$ from $<|H_{Fe}|> \approx 155$ kOe for $x=34.4$ to $ <|H_{Fe}|>\approx 60$ kOe for $x=55$.
The corresponding $<|H_{Fe}|>$-values obtained by means of the APM model also decrease linearly
with $x$ from $<|H_{Fe}|> \approx 105$ kOe to $ <|H_{Fe}|>\approx 40$ kOe, but at a slightly lower rate.  The measured values
of $<|H_{Fe}|>(x)$ show a non-linear behavior.  For the lowest V concentrated sample ($x=34.4$)
its value lies between those found for the FM and APM models.
For $x=40$ it is equal to the value calculated with the APM approach, while for higher $x$-values it
is smaller than the calculated values.
The observed quantitative agreement between the measured and the calculated values of
the hyperfine fields is not fully satisfying what can have different reasons. First of all, we should bear
in mind that in our present calculation only the Fermi contact term was taken into account, while other possible contributions like
dipolar and orbital ones, were neglected.
But we do expect that the most important influence on the computed $H_{hf}$-values could have a local ordering of magnetic
moments which mostly modify its valence terms. It is well seen on the dependencies of $H_{core}$ and $H_{val}$
contributions vs. magnetic moments
for FM model (Fig.~\ref{fig7}). As expected, $H_{core}$ is perfectly linear vs. $\mu$ with the expected
slope of about $10~T/\mu_B$ (both for Fe and V atoms whatever the sublattice), while for $H_{val}$ there is
no apparent correlation with the magnetic moment, since it strongly
depends on atomic and magnetic surroundings. In a real magnetic structure, valence contributions to $H_{hf}$ may
vary both in magnitude and in direction. The latter is not provided by the present calculations
when using simplified FM and APM models. It is worth noting that for Fe atoms the calculated $H_{hf}$ are
dominated by $H_{core}$ (best seen on A, D and E sites in Fig.~\ref{fig7}), which are fully correlated with local
magnetic moments. Since the calculated hyperfine fields overestimate the experimental data, one should expect
that the valence contributions, which are opposite in sign to the core contributions, may diminish this disagreement.
Hence, accounting for a more complex magnetic structure (with different local magnetic moments directions that mostly affect $H_{val}$)
could improve experiment versus theory agreement for the results on Fe atoms.

%We also expect that the relativistic corrections for not so heavy Fe and V nuclei are not so important when calculating hyperfine
%fields \cite{Dederichs90}.

On the other hand the correlation between the calculated average Fe-site hyperfine field and the
corresponding magnetic moment (per atom) - see Fig.~\ref{fig8} - agrees quite well with the measured one:
the slope of the experimentally found $<|H_{Fe}|>-<\mu_{exp}>$ line has the average value of
$\sim$ 19 T/$\mu_{B}$ if the magnetic moment is calculated per an atom in the unit cell
 (or 14.3 T/$\mu_{B}$ when the magnetic moment is determined per one Fe-atom in the unit cell
\cite{Cieslak09}), while that of the theoretically determined lines amounts to $\sim$14
T/$\mu_{B}$ for the FM ordering and to $\sim$27 T/$\mu_{B}$ for the APM one. In other words, the
experimentally found $<|H_{Fe}|>-<\mu_{exp}>$ line lies between the two theoretically
calculated ones. This can be taken as a further support for our approach that the magnetic
ordering in the studied system can be treated as a combination of FM- and APM-ordered regions.

%From the plots displayed in Fig.  7, one can estimate that the relative contribution of the FM domains
%ranges between ...  for $x=..$ and ...  for $x=..$, with the average of about 62\%.

Concerning the $^{51}$V fields, we have calculated the fields for four sublattices (B, C, D, and E) as
we can compare these quantities with experimentally determined ones for each of the five sublattices \cite{Dubiel10}.
In addition, the average values of the V-site field, $<|H_V|>$, were evaluated, too.
A comparison between the theoretical and experimental data, displayed in Fig.~\ref{fig6}b, gives an evidence
that the fields calculated for particular sites are underestimated by a factor of $\sim2$.
In consequence, the average field, namely the calculated $<|H_V|>$ decreases with $x$ at a lower
rate than the measured one. Using the same arguments as for the Fe-site hyperfine fields, we expect that $H_{val}$
are not sufficiently well evaluated within the FM and the APM models. Fig.~\ref{fig7}  shows that absolute values
of $H_{val}$ are slightly larger than the corresponding ones of $H_{core}$, resulting in relatively
small total hyperfine fields on V. Consequently, the calculated hyperfine fields became underetimated with respect
to the experimental ones.

\subsection{M\"ossbauer Spectra}

\begin{figure}[bt]
\includegraphics[width=.49\textwidth]{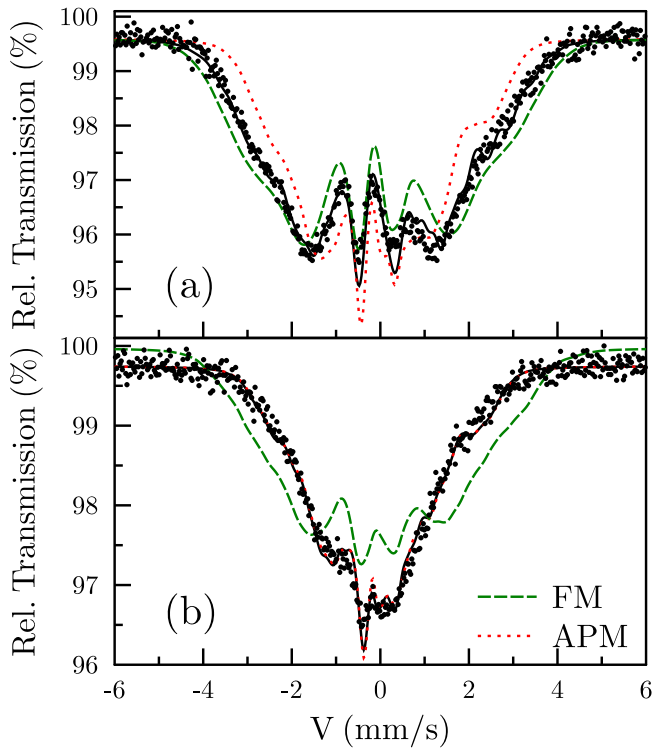}
\caption{(Online color)
$^{57}$Fe M\"ossbauer spectra recorded on the
(a) $\sigma$-Fe$_{65.6}$V$_{34.4}$ and
(b) $\sigma$-Fe$_{60.1}$V$_{39.9}$ sample at 4.2 K.
Dashed lines represent best fits to the data  calculated within the FM and the APM models.
}
\label{fig9}
\end{figure}

The calculations reported in this paper can be also used in the analysis of M\"ossbauer
spectra.  Towards this end we assumed the M\"ossbauer spectrum to be composed of five subspectra
correspondig to the $^{57}$Fe atoms located on the five inequivalent lattice sites.
The relative area of each subspectrum should be proportional to the number of Fe atoms occupying
a given site and it was assumed to be equal to that known from neutron diffraction experiments \cite{Cieslak08c}.
Each subspectrum represents a distribution of the hyperfine fields (hfd) with quadrupole splittings, $QS$, and relative
isomer shift-values, $IS$, identical to those determined from room temperature measurements and presented in
Ref. \onlinecite{Cieslak10a}. The hyperfine field distributions for each concentration were computed on the basis of
probability distributions and calculated $<|H_{Fe}|>(NN_{Fe})$ dependences obtained from FM and APM models.
As can be clearly seen in Fig.~\ref{fig9}, the M\"ossbauer spectrum recorded on the sample with $x=34.4$ could be
cuccessfuly fitted assuming the hfd to be weighted average of hfd obtained from the FM model
(weigt 58\%) and the APM one (42\%). For the sample with $x=40$ it was enough to take into accout the hfd
calculated for the APM model, only. For more concentrated samples it was not possible to get
any reasonably good fit on the basis of the calculated values.

\section{Conclusions}

The electronic structure calculations of ordered approximants of the $\sigma$-FeV phase were done
for border compositions of this phase, namely Fe$_{20}$V$_{10}$ and Fe$_{12}$V$_{18}$. The magnetic
properties, i.e. total and atomic magnetic moments, as well as Fe and V hyperfine fields were
calculated using two models for the magnetic moments alignment and compared to experimental values.
It was found that both $\mu(NN_{Fe})$ and $|H_{hf}|$ are zero below $NN_{crit}$, characteristic of each
lattice site, and increase more or less linearly for higher $NN_{Fe}$-values. The calculated $<\mu>$-values
overestimate the experimental data when using the FM model, and underestimate them in the case of the APM one.
This discrepancy was discussed in terms of the total energy difference as calculated for the FM and the APM models.
The analysis of the calculated Fermi contact terms of Fe and V hyperfine fields showed a crucial role of
valence contributions, being strongly dependent on local magnetic moments arrangements. The fact that only
simple magnetic structures were accounted for the KKR calculations, did not allow to expect fully
realistic determination of $H_{val}$. The enhancement of the $H_{val}$ contribution which either lowers
$|H_{hf}|$ on Fe atoms and increases this quantity on V atoms, would eventually lead to a better agreement with
the experimental data.
In spite of the aforementioned discrepancies between experimental and calculated values,
the almost linear dependences of $<|H_{Fe}|>(\mu)$, determined in $\sigma$-FeV phase for the FM and the APM models,
remains in satisfying agreement with measured one.

\begin{acknowledgments}
The results reported in this study were partly obtained within the project supported by the
  Ministry of Science and Higher Education, Warsaw (grant No. N N202 228837),
  by the European Communities under the contract of Association between EURATOM and IPPLM
within the framework of the European Fusion Development Agreement as well as
  by the FP7-NMP-2010-LARGE-4 Collaborative Project "AccMet" (No. 263206).
\end{acknowledgments}

\end{document}